\documentclass[preprint,showpacs,preprintnumbers,amsmath,amssymb,nofootinbib]{revtex4} 

\usepackage{graphicx}
\usepackage{dcolumn}
\usepackage{bm}
\usepackage{latexsym} 
\input amssym.def    
\input amssym.tex

\newcommand{\cL}{{\cal L}}

\newcommand{\nn}{\nonumber \\ }

\newcommand{\Tr}{\hbox{Tr}}
\newcommand{\dmudmuphi}{\partial_{\mu}\phi \partial^{\mu}\phi}

\def\undertext#1{\vtop{\hbox{#1}\kern 1pt \hrule}}
\def\ra{\rightarrow}

\def\lrb#1{\left(#1\right)}

\def\be{\begin{equation}}
\def\ee{\end{equation}}
\def\bea{\begin{eqnarray}}
\def\eea{\end{eqnarray}}

\def\beq{\begin{equation}}
\def\eeq{\end{equation}}
\def\bi{\begin{itemize}}
\def\ei{\end{itemize}}
\def\beqar{\begin{eqnarray}}
\def\eeqar{\end{eqnarray}}

\begin{document} 
\topmargin=0.1cm 
\preprint{MIT-CTP-3404} 
 
\title{The self-consistent bounce: an improved nucleation rate} 
 
\author{Yoav Bergner} 
\email{bergner@mit.edu} 
\affiliation{ 
Center for Theoretical Physics, 
Massachusetts Institute of Technology, 
77 Massachusetts Avenue, Cambridge MA 02139 
} 
\author{Lu\'{\i}s M. A. Bettencourt} 
\email{lmbett@lanl.gov} 
\homepage{http://www.mit.edu/~lmbett} 
\affiliation{ 
Los Alamos National Laboratory, MS B256, 
Los Alamos NM 87545 
} 
 
\date{\today} 
 
\begin{abstract} 
We generalize the standard computation of homogeneous nucleation 
theory at zero temperature to a scenario in which the bubble shape is determined 
self-consistently with its quantum fluctuations.  Studying two scalar 
models in 1+1 dimensions, we find the 
self-consistent bounce by employing a two-particle irreducible (2PI) 
effective action in imaginary time at the level of the Hartree approximation.  We 
thus obtain an effective single bounce action which determines the 
rate exponent.  We use collective 
coordinates to account for the translational invariance and the 
growth instability of the bubble and finally present a new 
nucleation rate prefactor.  We compare the results with those obtained 
using the standard 1-loop approximation and show that the 
self-consistent rate can differ by several orders of magnitude. 
\end{abstract} 
 
\pacs{\hfill} 
 
\maketitle 
 
\section{Introduction} 
During first order phase transitions, the macroscopic state of a  
system changes suddenly from the metastable phase (or false vacuum) to  
the stable phase (or true vacuum). If the transition is strongly first  
order it is expected to proceed by the spontaneous creation of  
spherically symmetric droplets -- or bubbles -- of the stable phase  
which, if large enough, can grow to consume the metastable phase.   
  
The modern theory of bubble nucleation is due principally to Langer  
\cite{Langer}, who was concerned with models of statistical physics  
and to Voloshin \cite{Voloshin} and Coleman and Callan  
\cite{Coleman,CallanColeman} who developed   
similar ideas in the context of relativistic (zero temperature)  
quantum field theory. The unification of Langer and Coleman's  
approach to finite temperature quantum field theory was later obtained  
by Affleck \cite{Affleck} and Linde \cite{Linde}, among others.    
  
Although the departure point for the theory of nucleation is  
completely general, in practice  the  
nucleation rate is computed almost without exception in the semiclassical approximation,  
including fluctuations only at most to 1-loop order.  Even for semirealistic  
models, these lowest order calculations can be computationally  
nontrivial \cite{Baacke}.  We quote the now well-known zero temperature  
result of Ref.~\cite{CallanColeman} that for $D$ spacetime dimensions,  
the nucleation rate per unit volume $\Gamma/V$ is given by  
 
\begin{eqnarray} 
\Gamma/V = A e^{-B}, \quad B = S_{cl}[\phi_b]/\hbar, \quad A = \left(\frac{B} {2 \pi}\right)^{D/2}~  \left| \frac{det' \left[ - \partial + V'' (\phi_b)\right]} {  det \left[ - \partial + V'' (\phi_+) \right]} \right|^{{-1/2}} 
\label{1loopRate} 
\end{eqnarray}  
where $S_{cl} = \int d^D x  (\partial_x \phi)^2/2 + V(\phi)$  is the  
classical Euclidean action of the ``bounce'' and $det'$ denotes, as usual, the  
fluctuation determinant with $D$ zero modes excluded. 
  
Nevertheless in certain instances, an improvement on  
the 1-loop computation is necessary, for example when the  
first order character of the transition is itself due to radiative  
corrections or to the presence of other fields,  
as in Higgs + gauge theories.  It would seem that, at least to lowest  
order again, these effects might be accounted for by substituting an  
effective potential for the classical one in the bounce computation  
or more generally employing an effective action  
\cite{Weinberg, Strumia}. But this is not a simple task as fluctuations exist  
over the bubble background, and this background may distort  
significantly the properties of the low lying spectrum relative to its  
perturbative form.    
  
One resolution to this problem is a different treatment of low and  
high energy fluctuations, for example in the   
context of a coarse-grained effective action  
\cite{Strumia,Wetterich,O'ConnorStephens}.  In such a procedure,  
however, much care must be   
taken to avoid double counting the effects of fluctuations, i.e. in  
their accounting for distortion of the bubble while also providing  
corrections to the background in which the bubble exists.   
Certain physical effects such as fluctuation backreaction on the  
classical bubble profile are usually absent from these  
calculations.  Finally, lattice Monte Carlo methods have also been  
used to study nucleation rates, including more recently for the case  
of radiatively induced transitions \cite{Alford, Moore}.  
  
In this paper we propose a self-consistently improved nucleation rate by  
including all of the above features in a semi-analytical  
computation \cite{Surig}.  We shall study two scalar    
models in 1+1 dimensions and include the self-coupling of  
fluctuations as well as their backreaction on the mean field  
nonperturbatively through the use of a two-particle irreducible (2PI) effective  
action formalism.  We have also used this technique recently to  
calculate the self-consistent quantum energies of  
topological defects \cite{kinkpaper}.    
  
The organization of the paper is as follows: in Sec.~\ref{ratereview}  
we review aspects of the constuction of the rate as defined in  
Eq.~(\ref{1loopRate}) as some of these will need to be modified later.  
We define our scalar models in Sec.~\ref{scaction} and give a short  
overview of the 2PI effective action formalism we employ.  We  
also define the quantum bounce    
in terms of coupled Euclidean spacetime equations for the mean field and the  
coincident two-point function.  In Sec.~\ref{screnorm} we discuss the  
renormalization structure of counterterms necessary to make these calculations   
finite.  We present numerical results for the  
quantum bounce profiles and corresponding effective action  
calculations in Sec.~\ref{numresults} and also discuss the spectrum of  
fluctuations in the self-consistent case vs at 1-loop order.  Finally  
in Sec.~\ref{newrate} we discuss how to extract the nucleation  
probability in the self-consistent case and propose a new nucleation  
rate.  A brief discussion of the results is offered in Sec.~\ref{discussion}
 
\section{Structure of the rate calculation} 
\label{ratereview}  
The 1-loop approximation to the rate  
calculation leading to Eq.~(\ref{1loopRate}) has several special features, which  
we now highlight in order to be able to contrast them with our results below.  
We will take $\hbar=1$ throughout and work in $D$-dimensional  
Euclidean spacetime, although later we shall specialize to $D=2$.  
  
The starting point of any rate calculation is the transition amplitude   
\begin{eqnarray}  
\langle \phi_{+} (x) \vert \phi_{-} (x') \rangle = {\cal N} \int d[ \phi ] e^{ - S [\phi]/\hbar}  = {\cal Z},  
\label{transamp} 
\end{eqnarray}  
with appropriate boundary conditions on the fields.  We are interested  
in a transition amplitude between the false and true vacua, i.e. the  
vacuum expectation value of the field is given by the true vacuum at  
some spacetime point, taken to be the origin   
$\langle \phi(x^{\mu}=0) \rangle= \phi_-$,  and in the false vacuum at  
spacetime infinity $\langle \phi(x^{\mu} \rightarrow \infty)  
\rangle = \phi_+$.  The normalization $\cal N$ ensures that
\beqar
\langle \phi_{+} (x) \vert \phi_{+} (x') \rangle = \langle \phi_{-}
(x) \vert \phi_{-} (x') \rangle = 1 .  
\eeqar
  
In the semiclassical approximation the path integral is evaluated by expanding around the field configurations that extremize the {\it classical }   
action, subject to these boundary conditions. The simplest such  
configuration is the classical bounce $\phi_{b}$. Well separated  
multi-bounce configurations also qualify and need to be summed over  
among the action extrema.    
  
The single bounce path integral can be evaluated semiclassically (i.e. to 1-loop order)  
by expanding the quantum field in linearized fluctuations around the  
classical bounce configuration   
\begin{equation}  
\phi = \phi_{b} + \sum_{n } a_{n} \psi_{n} .  
\end{equation}  
Then the (renormalized) path integral becomes  
\begin{eqnarray}  
{\cal Z} = {\cal N} A e^{-B}; \qquad   A=  \left|\frac{det \left[ -  
\partial + V'' (\phi_b)\right]} {det \left[-\partial + V'' (\phi_+)  
\right]}\right|^{-1/2}; \quad B = S_{cl}[\phi_b]  
\end{eqnarray}  
where we took the fluctuations to be eigenstates of the linearized  
operator $G^{-1}_{0} = - \partial + V'' (\phi)$.  Defined as such, the
fluctuations are non self-interacting and do not backreact on the profile
$\phi_{b}$ (i.e. neither the operator $G_0$ nor the equation of motion
solved by $\phi_b$ depends on the fluctuations $\psi_n$).    
  
Note that $\cal Z$ is dimensionless as it should be for a  
probability. There is a simple functional relation between $\cal Z$  
and the effective action $\Gamma_{\rm eff}$. In the absence    
of external sources   
\begin{eqnarray}  
{\cal Z} = e^{-\Gamma_{\rm eff}}  
\end{eqnarray}  
and the 1-loop effective action takes the well known form $\Gamma_{\rm  
eff} = B - \frac12 \ln A$.   
  
Given the transition amplitude due to one bounce, it is  
straightforward to generalize to any number of bounces. This is the  
sum of the 1-bounce contribution plus    
the 2-bounce contribution, which for well separated configurations becomes the product   
of two single bounces, etc. The result of resumming this series is the  
exponentiation of the single bounce amplitude i.e.  
\begin{eqnarray}  
Z = \exp[ e^{-B}  A]  = \exp[ e^{-\Gamma_{\rm eff}}]    
\end{eqnarray}  
  
The nucleation rate (per unit volume) has dimensions of inverse  
spacetime volume and can be obtained directly from the multi-bounce amplitude as    
\begin{eqnarray}  
\Gamma/V =  2\  \hbox{Im}\lrb{{1 \over \Omega} \ln   
{\cal Z}},
\label{imagpart}
\end{eqnarray}  
where $\Omega$ is the spacetime volume. (We regret the confusing  
notation which arises since the capital letter $\Gamma$ serves as the  
conventional standard for both the rate and the effective  
action.  We use $\Gamma_{\rm eff}$ where necessary to avoid ambiguity.)  
   
 To see that this result is finite in the infinite volume limit, we  
must analyze the determinant ratio in the prefactor   
$A$ in greater detail. To 1-loop order, it is well known that the  
fluctuation spectrum has one negative    
eigenvalue (an imaginary frequency) and $D$ zero eigenvalues.   
The former keeps track  of the instability of the false vacuum. In a  
theory without backreaction this instability is connected with a    
nonconservation of a probability current \cite{CallanColeman,Affleck},  
a mode that will grow unchecked in the large spacetime    
volume. The $D$ zero modes are proportional to the $D$ variations  
$\partial_{\vec x} \phi_{b}$,    
and represent the translational invariance of the ``center'' of the classical bounce.  
  
Each eigenvalue in the determinant $A$ results from a Gaussian integration   
\begin{eqnarray}  
\frac{1}{\omega} = \frac{1}{2 \pi} \int d \psi e^{- \psi {\omega^2_i \over 2}  \psi}.  
\end{eqnarray}    
 This integral is clearly divergent if $\omega^{2}$ is negative, but  
 it is rendered sensible   
 by deformation of the contour into the complex plane (in effect taking   
 $\omega^{2} \rightarrow |\omega^2|$).   
 Thus the amplitude $\cal Z$ is imaginary, reflecting the fluctuation  
 instability. (As the nucleation--or decay--rate is determined by the imaginary  
 part of the transition amplitude, one simply excises the factor of $i$).   
  
In the case $\omega \rightarrow 0$ the corresponding integral is of  
course not Gaussian at all. It   
becomes, up to a Jacobian factor, the volume and diverges with it.   
This is clear from the change of variables   
\begin{eqnarray}  
[d \phi ]i = J dz, \qquad J = \left[  \int d^D x  \left( \frac{  
\partial \phi(x,z)}{\partial x}  \right)^2  \right]^{1/2},  
\eea  
\bea  
\lim_{\omega\rightarrow 0} \frac{1}{\omega} = \sqrt{\frac{1}{2 \pi}} J  \int d z = \frac{J}{\sqrt{2 \pi}}  
\Omega^{1/D},  
\end{eqnarray}  
with the linear volume $\Omega^{1/D} \rightarrow \infty$. What we have  
done is exchange the functional integration over the field   
direction $[d \phi]$ for an integration over a collective coordinate  
$z$.   
The zero modes thus give rise to an infrared divergence (related to  
the limit of infinite spacetime volume) in the determinant prefactor. This factor of volume is  
happily canceled as the probability amplitude, taken per unit spacetime  
volume, becomes the desired rate per volume of Eq.~(\ref{1loopRate}).  Namely   
\begin{eqnarray}  
\frac{1}{\Omega}\left|\frac{det\left[-\partial + U''  
(\phi_b)\right]}{det\left[-\partial + U'' (\phi_+) \right]}  
\right|^{-1/2} \ra    
  \lrb{\frac{J}{\sqrt{2 \pi}}}^D \left|\frac{det' \left[ - \partial + U'' (\phi_b)\right]} {  det \left[ - \partial + U'' (\phi_+) \right]} \right|^{-1/2}.  
\end{eqnarray}  
 (In a finite volume  
application, it is quite reasonable that the nucleation rate grows as  
a function of volume.)  
  
There are two distinct ways in which this standard calculation is   
approximate. The first is the computation of the single-bounce effective  
action. The semi-classical or 1-loop approximation is a lowest   
order calculation of the effects of fluctuations.  It completely   
neglects both fluctuation self-interactions and their backreaction on   
the classical field profile.  Because barrier nucleation is a  
Boltzmann-suppressed process, exponentially sensitive to corrections,  
the inclusion  of these effects may lead to large quantitative  
differences. A computation of the nucleation rate that includes these  
effects is the subject of the present paper.  
  
The second approximation involved in the rate calculation is the ``dilute   
instanton approximation''  \cite{CallanColeman,tHooft} for the multi-bounce configurations.   
This is in principle justified so long as the nucleation rate   
per volume is sufficiently small. It will not be a good approximation  
though in cases where    
nucleation may be enhanced by earlier bubbles or by the presence of   
other nonperturbative fluctuations over the homogeneous background.   
Although some of these situations could be addressed with the techniques   
developed here we will not explore them in the present manuscript.  
We will thus continue to assume the dilute instanton approximation   
as an unaltered ingredient of the rate computation.

\section{2PI Effective Action} 
\label{scaction} 
The traditional starting point for the calculation of the single 
bounce is the    
classical action.  We refer the reader to the Ref.~\cite{Coleman} 
for that formalism and do not summarize it extensively here.  Suffice it to say that the 
classical bounce is an $O(D)$-symmetric solution to the Euclidean 
field equations in $D$ spacetime dimensions satisfying certain 
boundary conditions, in particular that the field assumes a true vacuum expectation 
value in a local region around the origin and a false vacuum 
expectation value far away and to infinity.  In our development of the 
self-consistent bounce, we begin instead with the two-particle irreducible (2PI)   
effective action for the field $\phi$ and the two-point function 
$G(x,x')$ developed in the relativistic context by Cornwall, Jackiw 
and Tomboulis (CJT) \cite{2PI}.  
 
The effective action functional is given by 
\beqar 
\Gamma(\phi,G)=I(\phi) + \frac12 i \Tr \ln D_0 G^{-1}  
	+\frac12 i\Tr D^{-1}(\phi) G + \Gamma_2 (\phi,G) + \hbox{const} 
\label{effaction} 
\eeqar 
In this notation we have the classical action 
\beqar 
I(\phi) = \int d^2 x  \cL[\phi], \qquad \cL[\phi] = \frac12 \dmudmuphi - V(\phi) 
\eeqar 
and the operator 
\beqar 
i D^{-1}(x,x';\phi) &=& \frac{\delta^2 I}{\delta \phi(x) \delta 
\phi(x')} . 
\eeqar 
$D_0$ is the free propagator. 
Finally $\Gamma_2$ sums the 2PI vacuum to vacuum diagrams which furnish 
an expansion in the number of loops or, equivalently, in powers of 
$\hbar$ ($\hbar=c=1$ below and throughout).    
Stationarizing the action leads to the equations: 
\beqar 
\label{fieldeq} 
\frac{\delta I}{\delta \phi} + \frac12i G \frac{ \delta 
D^{-1}}{\delta \phi} + \frac{\delta \Gamma_2}{\delta \phi} &=& 0 \\ 
- \frac12i G^{-1} + \frac12i D^{-1} + \frac{\delta \Gamma_2}{\delta G} 
&=& 0 
\label{propeq} 
\eeqar 
We shall consider two symmetry-breaking forms for the scalar potential 
\beqar 
\label{p4model}
V_4 (\phi) &=& \frac14 \lambda \left[ (\phi^2 - v^2)^2 - 
\frac{2\epsilon}{\lambda v}(\phi+v) \right] \qquad v^2 = 
\frac{\mu^2}{\lambda}  
\\ 
V_6 (\phi) &=& \frac12 \lambda \phi^2 \left[ (\phi^2 - a^2)^2 - 
\frac{2\epsilon}{\lambda a^2} \right] \qquad a^2 = \frac{\mu}{\sqrt{\lambda}} 
\label{p6model}
\eeqar 
These potentials represent more or less generic models with true and 
false vacua at the classical level and have been cast in the form that the 
potential difference between the true and false vacua is $\epsilon$. 
The symmetry is explicitly broken in the quartic model while a $\phi 
\ra -\phi$ discrete symmetry still exists in the sextic model (where 
the false vacuum is the symmetric minimum $\phi=0$).  The quantization 
of solitons in (1+1)-dimensional models of precisely these types with 
$\epsilon=0$ were originally 
examined in Ref.~\cite{Lohe}.  
 
For these models we obtain the following equations: for $V_4$, 
\beqar 
&& \Box \phi + \frac12 \lambda 
\left[ 2 \phi (\phi^2 - v^2) - \frac{\epsilon}{\lambda v} \right] 
+ 3 \lambda \phi G(x,x) - \frac{\delta \Gamma_2}{\delta \phi} = 0 
\\ 
&& i G^{-1}(x,x') + \left(\Box_x - \lambda v^2 + 3 \lambda 
\phi^2 \right)\delta^2(x-x')  - 2\frac{\delta \Gamma_2}{\delta G} 
= 0 
\eeqar 
and for $V_6$, 
\beqar 
&& \Box \phi + \lambda \phi 
\left[ (\phi^2 - a^2)^2 +2\phi^2(\phi^2 - a^2) - \frac{2\epsilon}{\lambda a^2} 
\right] - 6 \lambda \left( 2 a^2 \phi - 5 \phi^3 \right) G(x,x) - \frac{\delta \Gamma_2}{\delta \phi} = 0 \nn 
&& \phantom{1=1} \\ 
&& i G^{-1}(x,x') + \left[\Box_x + \lambda (a^4-\frac{2\epsilon}{\lambda a^2})-12 \lambda a^2 \phi^2 + 15 \lambda 
\phi^4 \right]\delta^2(x-x') - 2\frac{\delta \Gamma_2}{\delta G} 
= 0 \nn 
&& \phantom{1 = 1} 
\eeqar 
 
Up to this point we have made no approximations, but we have left 
$\Gamma_2$ as the sum of 2PI vacuum-to-vacuum diagrams with interaction 
vertices determined by the shifted Lagrangian and lines representing 
the full propagator $G$.  For our models, we can summarize the 
interactions in Table~\ref{vertices}.  
\begin{table} 
\begin{tabular}{lcccc} 
\hline \hline 
\emph{vertex} && $I_4$ && $I_6$ \\ \hline 
cubic && $-\lambda \phi$ && $-2  \lambda (5 \phi^3 - 2 a^2 \phi)$ \\ 
quartic && $-\frac{\lambda}{4}$ && $-\frac{\lambda}{2} (15\phi^2 - 2 
a^2)$ \\ 
quintic && && $-3\lambda \phi$ \\ 
sextic && && $-\frac{\lambda}{2}$ \\
\hline \hline
\end{tabular} 
\caption{Interaction vertices used in the construction of the 2PI 
effective action for two models} 
\label{vertices} 
\end{table} 
 
\begin{figure} 
\includegraphics[width=2in]{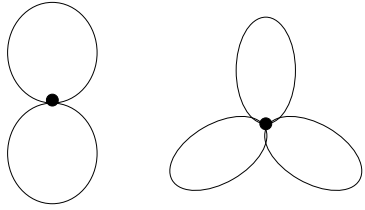} 
\caption{\label{diagrams} Vacuum bubble diagrams in the Hartree approximation for the  
2PI effective action of our models.  For the classical action $V_4$, 
only the left diagram contributes. Lines denote the full propagator $G$ in the nontrivial $\phi$ background.} 
\end{figure} 
 
It is of practical necessity in the computation to select only certain 2PI diagrams, and 
such a choice will correspond to a truncation in the Dyson-Schwinger 
hierarchy of equations for the correlation functions of the model.  It is 
straightforward to write down the full evolution equations in the 
Hartree approximation, which amounts to including only the 2PI 
``bubble'' diagrams shown in Fig.~\ref{diagrams}.  In $O(N)$ models,
this truncation is similar to the leading-order large-$N$
approximation.  While the Hartree approximation is not a controlled,
systematic expansion, it is understood to be 
equivalent to a Gaussian variational {\em ansatz} in the Schr\"odinger 
functional formalism \cite{Schroedinger} and results
in Hamiltonian dynamics \cite{Hamiltonian}.  Furthermore, a study of the quantum
energy of solitons showed good agreement between the Hartree values
and ``exact'' lattice Monte Carlo results \cite{kinkpaper, salle}.  Proceeding with the
approximation, we find
\beqar 
\Gamma_2^{V_4} &=& -\frac{3}{4}\lambda \int d^2 x G(x,x)^2  
\\ 
\Gamma_2^{V_6} &=& -\frac{3}{2}\lambda \int d^2 x (15 \phi(x)^2 - 2 a^2)  
G(x,x)^2 -\frac{15 }{2} \lambda \int d^2 x G(x,x)^3 \ . 
\eeqar 
so we obtain  
for $V_4$ (after inverting the $G$ equation), 
\beqar 
\label{phieq4} 
&&\Box \phi + \frac12 \lambda 
\left[ 2 \phi (\phi^2 - v^2) - \frac{\epsilon}{\lambda v} \right] 
+ 3 \lambda \phi G(x,x) = 0 \\ 
&&  \left[\Box_x - \lambda v^2 + 3 \lambda 
\phi^2 +3\lambda G(x,x)\right] G(x,x') 
= -i \delta^2(x-x') 
\label{Geq4} 
\eeqar 
and for $V_6$, 
\beqar 
&& \left[\Box + \lambda \left[ (\phi^2 - a^2)^2 +2\phi^2(\phi^2 - a^2) - \frac{2\epsilon}{v} \right] \right.\nn 
\label{phieq6} 
&& \qquad - \left. 6 \lambda \left( 2 a^2 - 5 \phi^2 \right) 
G(x,x) + 45 \lambda G(x,x)^2 \right]\phi = 0 \\ 
&& \left[\Box_x + \lambda 
(a^4-\frac{2\epsilon}{\lambda a^2})-12 \lambda a^2 \phi^2 + 15 \lambda 
\phi^4 \right. \nn 
&& \qquad \left. + 6 \lambda (15\phi^2-2a^2) G(x,x) + 45\lambda G(x,x)^2 \right] G(x,x')= -i \delta^2(x-x') \nn 
\label{Geq6}\phantom{1} 
\eeqar 
This formal procedure has absorbed the approximate quantum dynamics of 
the scalar field $\phi$ into a truncated set of equations to be solved 
simultaneously.  The two-point function $G(x,x')$ now appears 
explicitly as a ``sort of'' Green's function, albeit of a nonlinear 
operator which depends on its value at coincident spacetime points. 
Typically we would proceed by decomposing  $G(x,x')$ in a mode basis 
(related to the decomposition of the field operator of which it is a 
correlation function).  The mode functions and the field then satisfy 
a tower of partial differential equations which can be stepped forward 
in time on a computer.  This would be a prescription for the nonequilibrium 
evolution given Cauchy-type initial conditions \cite{stepbeyond}.   
 
\subsection{Imaginary Time} 
We shall alternately consider attempting to solve 
Eqs.~(\ref{phieq4})-(\ref{Geq6}) in imaginary time.  Passing to 
Euclidean spacetime and  
framing the partial differential equations as elliptical equations 
instead of hyperbolic ones does not in and of itself simplify the problem 
unless we can exploit special symmetries.  This is precisely what is 
done in the case of the bounce, where as a consequence of Lorentz 
invariance we seek a rotationally invariant classical solution, 
i.e. we need solve only for radial functions.  We can now see why at least 
at the level of the Hartree approximation, we can attempt a similar 
sleight of hand on the two-point function and the field 
simultaneously.  Wherever the two-point function appears in the 
evolution operators, it appears with coincident spacetime arguments. 
Therefore, we can seek solutions to Eqs.~(\ref{phieq4})-(\ref{Geq6}) 
in imaginary time in which both $\phi(x)$ and $G(x,x)$ 
depend only on the radial coordinate in Euclidean spacetime.  
 
To wit, the imaginary time formulation of the problem entails, 
\beqar  
t \rightarrow i\tau , \qquad  \Box \rightarrow \Box_E = -\nabla^2 
\eeqar 
or going to polar coordinates 
\beqar 
\Box_E = - \frac{1}{\rho}\frac{\partial}{\partial \rho}\left(\rho 
\frac{\partial}{\partial\rho} \right) - 
\frac{1}{\rho^2}\frac{\partial^2}{\partial \theta^2}  
\eeqar 
We seek a solution to the field VEV and two-point function equations 
where the field VEV and the coincident two-point function are 
purely radial, i.e. 
$$ 
\phi=\phi(\rho), \qquad \qquad G(\rho,\theta;\rho,\theta)=G(\rho) 
$$ 
while in general there remains $\theta$-dependence in $G(\rho,\theta;\rho',\theta')$. 
It is consistent with these constraints to separate variables in which case $G$ takes the form 
\beq 
G(\rho,\theta;\rho',\theta') = \sum_{n,l} a_{nl} R_{nl}(\rho) R_{nl}(\rho') e^{i l (\theta-\theta')} 
\eeq 
The $\phi$ equations (\ref{phieq4}) and  (\ref{phieq6}) become 
\beqar 
\label{phieuclidean} 
&&\Box_E \phi + \frac12 \lambda 
\left[ 2 \phi (\phi^2 - v^2) - \frac{\epsilon}{\lambda v} \right] 
+ 3 \lambda \phi G(\rho) = 0 \\ 
&& \left[\Box_E + \lambda \left( (\phi^2 - a^2)^2 +2\phi^2(\phi^2 - a^2) - \frac{2\epsilon}{\lambda a^2} \right) \right.\nn 
&& \qquad - \left. 6 \lambda \left( 2 a^2 - 5 \phi^2 \right) 
G(\rho) + 45 \lambda G(\rho)^2 \right]\phi(\rho) = 0 
\eeqar 
and the equations 
\beqar 
&& \left[ \Box_{\rm E} +\chi(\rho) \right]G(\rho,\theta;\rho',\theta') =\frac{1}{\rho}\delta(\rho-\rho')\delta(\theta-\theta'), 
\label{Geuclidean} \\ 
&& \chi_4(\rho) = -m^2 + 3 \lambda \phi^2 +3\lambda G(\rho). \nn 
&& \chi_6(\rho) = \lambda (a^4-\frac{2\epsilon}{\lambda a^2}) 
-12 \lambda a^2 \phi^2 + 15 \lambda\phi^4  
+ 6\lambda (15\phi^2-2a^2) G(\rho) +45\lambda G(\rho)^2. \nonumber 
\eeqar 
lead to the following ``nonlinear eigenvalue'' problem for the radial modes 
\beq 
\label{eigenvalue} 
\left[ - \frac{1}{\rho}\frac{\partial}{\partial \rho}\left(\rho 
\frac{\partial}{\partial\rho} \right) + 
\frac{l^2}{\rho^2} +\chi(\rho) \right] R_{nl} = \omega^2_{nl} R_{nl}. 
\eeq 
with the identification 
\be 
a_{nl}^{-1} = \omega^2_{nl}, \qquad 
G(\rho) = \sum_{n,l} \frac{R^2_{nl}(\rho)}{ \omega^2_{nl}} 
\ee 
provided the orthonormality and integrability condition holds for the $R_{nl}(\rho)$, 
\beqar 
2 \pi \int_0^{\infty} \rho d \rho R_{nl} (\rho) R_{mk} (\rho) = \delta_{nm,kl} 
\eeqar 
 
We must also impose boundary conditions at the origin $\rho=0$, where 
for analyticity we require that $G$ have zero derivative, and at 
large values of $\rho$ where $G$ should approach its value in the 
false vacuum.  We impose the first condition on the radial mode 
functions $R_{nl}(\rho)$ whereby for $l=0$ the functions must have 
zero derivative at the origin and for $l \ne 0$ they must vanish.  The 
condition at large $\rho$ can be made consistent with Dirichlet 
boundary conditions.  This latter condition is especially suitable if the renormalized 
value of $G$ is fixed to be zero in the false vacuum.  We 
shall discuss this renormalization condition in the following section.

\section{Renormalization of Self-consistent Effective Action} 
\label{screnorm} 
Although any scalar model in 1+1 dimensions is super-renormalizable, 
details of the renormalization differ slightly in the self-consistent
effective action from the usual loop expansion. Moreover numerical
implementation of renormalization schemes comes with its own
subtleties.  Here we detail the 
procedure for a simplified $\phi^4$ model, i.e. with no symmetry breaking
parameter $\epsilon$, in two steps.  We first demonstrate 
the analytical cancellation of divergences assuming a homogeneous 
background.  This may be considered a review of the renormalization 
of the self-consistent effective potential as worked out previously, 
for example in Ref.~\cite{ACP}.  
Next we describe the changes which occur for  the inhomogeneous case 
of interest and for a finite volume with a discrete spectrum--as is the case in any numerical analysis. 
 
For homogeneous fields we consider the effective potential and pass to 
Euclidean momentum space for the translationally invariant two-point function, i.e. 
\be 
V(\phi,G)\int d^2 x = \Gamma(\phi,G) , 
\ee 
\be 
G(x,x') = \int \frac{d^2 k}{(2\pi)^2} e^{-ik(x-x')}G(k) \equiv \int_k 
e^{-ik(x-x')}G(k) , 
\ee 
such that 
\beqar 
\label{Veffdef} 
V(\phi,G) &=& V_{cl}(\phi) + \frac{1}{2}\int_k \, \ln D_0 G^{-1} + \frac{1}{2}\int_k 
\, \left[ D^{-1} G -1 \right] + \frac{3\lambda}{4}\Bigl[\int_k \, 
G(k) \Bigr]^2  
\eeqar 
Now setting $\frac{\delta V}{\delta G} = 0$, we have 
\beqar 
G^{-1}(k) =  D^{-1}(k) + 3 \lambda \int_k \, G(k)  
\label{gap} 
\eeqar 
and taking as an {\em ansatz} $G(k) = \frac{1}{k^2+M^2}$, we obtain a 
gap equation for $M^2$: 
\beqar 
M^2 = -m^2 + 3\lambda \phi^2 +3 \lambda \int \frac{d^2k}{(2\pi)^2} 
\frac{1}{k^2 +M(\phi)^2} 
\label{gapM} 
\eeqar 
The integral in the gap equation is logarithmically divergent, hence $M^2$ is so 
far undefined.  We now demonstrate the counterterm procedure 
which renormalizes this equation and also renormalizes the effective 
potential. 
We first rewrite the equation as 
\beqar 
M^2 = -m^2 + 3\lambda \phi^2 +3 \lambda \lrb{\int \frac{d^2k}{(2\pi)^2} 
\frac{1}{k^2 +M(\phi)^2} - I} + 3\lambda I 
\label{massgap} 
\eeqar 
and impose a momentum cutoff $\Lambda$ with 
\be 
I = \int^\Lambda  \frac{d^2k}{(2\pi)^2} G_0(k) \equiv \int^\Lambda 
\frac{d^2k}{(2\pi)^2} \frac{1}{k^2 + \mu_0^2} = 
\frac{1}{4\pi}\ln\frac{\Lambda^2}{\mu_0^2} . 
\ee 
This implements the regularization of the integrals in Eq.~(\ref{massgap}). 
The new scale $\mu_0$ is necessary to prevent an infrared logarithmic
divergence in the infinite volume limit.  The renormalization
procedure is now carried out by identifying the bare   
parameter $m^2$ as a function of the cutoffs and defining a
renormalized (physical) mass by
\be 
-m_R^2 = -m^2 + \delta m_1^2 = -m^2 + 3\lambda I 
\ee 
In terms of the renormalized mass, the gap equation is finite 
\beqar 
M^2 &=& -m_R^2 + 3\lambda \phi^2 + \frac{3 \lambda}{4\pi} 
\ln\frac{\mu_0^2}{M^2} \\ 
&=& -m_R^2 + 3\lambda \phi^2 + 3 \lambda G_f(M). \label{renormgap} 
\eeqar 
and we can see that by a choice of renormalization scale ($\mu_0^2 =
2m_R^2$), the
self-consistent fluctuation mass in the classical vacuum $\phi_0^2 = m_R^2/\lambda$
can be fixed to its ``classical'' value $M^2=2m_R^2$.  We shall consider
$\mu_0$ to be a physical scale, a function of the physical mass and
couplings.  
 
We now express the Hartree resummed effective potential in terms 
of its contributions at different orders, following Ref.~\cite{ACP}. 
\beqar 
V_{eff}= V_0 + V_1 + V_2 
\eeqar 
Substituting Eq.~(\ref{gap}) into (\ref{Veffdef}), we now have 
\beqar 
V_0 &=& V_{cl} = -\frac12m^2\phi^2 + \frac{\lambda}{4}\phi^4 
\nn 
V_1 &=&  \frac{1}{2} \Tr \ln D_0 G^{-1} =  \frac{1}{2}\int \frac{d^2k}{(2\pi)^2} \ln \left[\frac{k^2+M^2}{k^2+\mu_0^2} 
  \right] \nn 
V_2 &=& -\frac{3\lambda}{4} G(x,x)^2 = -\frac{3\lambda}{4} 
    \left(\int \frac{d^2k}{(2\pi)^2} \frac{1}{k^2+M^2} \right)^2 \ , 
\eeqar 
Note that as a simplification this classical potential has no symmetry breaking term and no 
overall constant.  The constant has no physical significance, but we 
will in any case make the shift in the potential explicit below by 
subtracting the value at a specified vacuum. The symmetry breaking 
coupling introduces no new infinities and requires no renormalization.  
 
We use (\ref{gapM}) to write  
\beq 
G(x,x) = \frac{1}{3\lambda} \lrb{M^2-m^2-3\lambda\phi^2} , 
\eeq 
and 
\be 
V_0+V_2= -\frac{M^4}{12\lambda}-\frac12\lambda\phi^4 
+ \frac{1}{6\lambda}M^2 m^2 + \frac12 M^2 \phi^2  , 
\ee 
(where an additive constant has again been removed).  We rewrite in terms 
of renormalized parameters, 
\be 
-m^2 = -m_R^2 - 3\lambda I, \qquad I=G_0(x,x)=\frac{1}{4\pi} \ln\frac{\Lambda^2}{\mu_0^2} 
\ee 
and use the renormalized gap equation (\ref{renormgap}) to write 
\be 
\frac{1}{2}M^2 G_f(M) = \frac{1}{6\lambda} M^4 +  \frac{M^2 
  m_R^2}{6\lambda} -\frac12 M^2\phi^2 , 
\ee 
such that 
\be 
(V_0+V_2)^R= \frac{M^4}{12\lambda}-\frac12\lambda\phi^4 - 
\frac{1}{2}M^2 G_f(M) - \frac{1}{2} M^2 I . 
\ee 
Finally, the one-loop contribution to the potential comes to 
\bea 
V_1 &=& \frac{1}{2}\int \frac{d^2k}{(2\pi)^2} \ln \left[\frac{k^2+M^2}{k^2+\mu_0^2} 
  \right]  \\  
&=& \frac{1}{8\pi}(M^2-\mu_0^2)\lrb{1+\ln\frac{\mu_0^2}{M^2}} + 
\frac{1}{2} (M^2-\mu_0^2) I . 
\eea 
After shifting the combined tree-level and 2PI pieces of the 
effective potential by their contribution in the classical vacuum 
$\phi_0^2 = m_R^2/\lambda$ (where the renormalization scale is set),  
the renormalized total is 
\be 
V_{eff} = -\frac12 \lambda(\phi^4-\phi_0^4) + \frac{M^4-\mu_0^4}{12\lambda}+ \frac{1}{8\pi}(M^2-\mu_0^2), 
\ee 
which is finite.   
 
The generalization of the effective potential to the 
effective Euclidean action for inhomogeneous mean field background now follows. 
We no longer have a simple gap equation, however the classical field $\chi$ 
defined in Eq.~(\ref{Geuclidean}) 
now replaces the $M^2$ in the above analysis.  The counterterm 
renormalization goes through and renders the tree level and 2PI 
components as a finite piece, expressed in terms of renormalized 
parameters, plus a counterterm: 
\bea 
\label{finitegamma} 
\Gamma_{0+2}  = \Gamma^f_{0+2} - \frac12 I  \int dx \lrb{\chi(x)-\chi_0}  
\eea 
where 
\bea 
\Gamma^f_{0+2} &=& \int d^2x \lrb{\frac12(\nabla\phi)^2+ \frac{\chi^2(x)}{12\lambda}-\frac12\lambda\phi^4 - 
\frac{1}{2}\chi(x) 
G_f(x)-\frac{\chi_0^2}{12\lambda}+\frac12\lambda\phi_0^4} . 
\eea 
We have again chosen the mass scale of the counterterm to set $G_f(\phi_0)= 0$, so 
\bea 
\chi(x) &=& -m_R^2 + 3\lambda \phi^2(x) + 3 \lambda G_f(x),  \nn 
\chi_0 &=& -m_R^2 + 3\lambda \phi_0^2 , 
\label{chi} 
\eea 
and we could have equivalently written 
\bea 
\Gamma^f_{0+2} &=& \int d^2x \lrb{V_{cl}(\phi)-V_{cl}(\phi_0)-\frac{3\lambda}{4} G_f(x,x)^2 
} , 
\eea 
with $V_{cl}$ expressed in terms of renormalized (finite) parameters. 
Finally, the one-loop order contribution in the inhomogeneous case is 
obtained from the eigenvalues of  Eq.~(\ref{eigenvalue}). 
\bea 
\label{oneloopterm} 
\Gamma_1 &=& \frac12 \sum \!\!\!\!\!\!\! \int \ln 
\frac{\omega^2_{nl}}{(\omega_0^2)_{nl}} , 
\eea 
by which we mean that $\omega^2_{nl}$ are the eigenvalues in the 
presence of the spatially varying field $\chi(\rho)$ and 
$(\omega_0^2)_{nl}$ are the eigenvalues in the vacuum defined by 
$\chi_0$.  This term is still logarithmically divergent, but this 
divergence is cancelled by the second term in  Eq.~(\ref{finitegamma}) 
with  
\be 
I \equiv G_0(\rho) = \sum_{n,l} \frac{(R_0^2)_{nl}(\rho)}{(\omega_0^2)_{nl}}. 
\label{cterm} 
\ee 

The renormalization of the $\phi^6$ model follows along the same lines as
above, although it is somewhat more cumbersome.  In that case there occurs a
renormalization of both the quadratic and quartic couplings.  Both
divergences have the same origin in the logarithmic loop integral of
a gap equation analogous to Eq.~(\ref{gapM}), i.e. there is still only
one renormalization scale.

There remains a subtlety in the practical computation of 
this one-loop term (\ref{oneloopterm}), which has been studied in the context of quantum 
corrections to static solitons \cite{kink}.  In that case, the 
quantity of interest is the sum over zero-point energies, or more 
precisely the difference of this sum in the presence of a soliton from this sum in 
the vacuum.  Rebhan and van Nieuwenhuizen have pointed 
out that in the lattice discretized sum (for example in a finite 
volume) care must be taken in implementing a frequency cutoff for the 
sum such that the same number of modes are included in the vacuum sum 
and in the nontrivial sum (ambiguity arises because of both the 
presence of zero modes at the one-loop level and the phase shifts of 
the ``continuum'' states).  We shall refer to this as a frequency 
cutoff with mode parity (FCMP) instead of a mode number cutoff as in Ref.~\cite{kink}.  The analysis here of a spherically 
symmetric background in higher dimensions must follow the same 
prescription, and moreover the FCMP must be implemented separately for 
each angular quantum number.  In other words, there is one frequency cutoff 
defined by the lattice spacing, but this leads to the inclusion of a 
different number of radial modes of each angular index included in the sum. 
 
\section{The single bounce effective action} 
\label{numresults} 
We have implemented the procedure described in the previous sections 
on a desktop workstation.  Namely, we compute the vacuum counterterm 
(\ref{cterm}) by solving the radial eigenvalue problem 
(\ref{eigenvalue}) in the false vacuum of each model.  We then 
introduce a nontrivial mean field bubble of true vacuum (using the 
semi-classical bounce as an initial guess\footnote[1]{The initial guess 
is highly arbitrary and our use of the semi-classical bounce does not for 
example preclude the application of this method to a theory with a 
radiatively induced phase transition.  The nature of the algorithm 
allows our initial guess to relax into the correct bounce profile 
provided the boundary conditions, e.g. the approximate symmetry 
breaking vacuum expectation values, are initially satisfied.}) and solve the renormalized 
versions of 
Eqs.~(\ref{phieuclidean})-(\ref{Geuclidean}) simultaneously by successive 
iteration.  We use a standard relaxation routine \cite{Nrecipes} to solve Eq.~(\ref{phieuclidean}),  
together with a set of LAPACK packages \cite{LAPACK} to solve the 
eigenvalue problem Eq.(\ref{eigenvalue}).  At each step,  the 
fluctuation eigenproblem is solved to produce a new intermediate 
$G_{\rm int}(x)$, which we combine with the old value to produce a new 
trial $G_{\rm new} = (1-\gamma) G_{\rm old} + \gamma G_{\rm int}$, 
until convergence in $\varphi$ and $G(x)$ is reached, up to a specified precision.  
The adjustable parameter $0 < \gamma < 1$ controls the size of the 
update and can be adapted to optimize convergence.   
 
This numerical procedure will find the correct shape of 
the mean-field profile and the coincident two-point function if the
radius is constrained.  Unconstrained, however, it may fail to find the
correct radius of the bubble, given the initial  
guess, since the stationary point of the action is a saddle 
point, not a minimum.  (In other words, the successive iteration will
not flow towards the saddle point.)  The self-consistent bounce typically has a 
smaller radius than the classical bounce.  We effectively constrain
the radius first at its classical value and then vary the radius to
find the actual saddle point.
 
The action is computed at each iterative step.  It should be noted 
that the very first step of the procedure amounts to computing the 
fluctuation contribution to the action in the presence of the 
classical bounce without backreaction on the mean field or 
self-coupling of the fluctuation modes.  In other words it is the 
standard 1-loop correction.  Thus we are able to compare the 
value of the Euclidean action at the classical, 1-loop and self-consistent 
levels\footnote[2]{In the one loop case we excise the zero modes and take the  
absolute value of the negative mode, as usual.}.
 
\begin{figure} 
\vspace{0.5in} 
\includegraphics[width=5in]{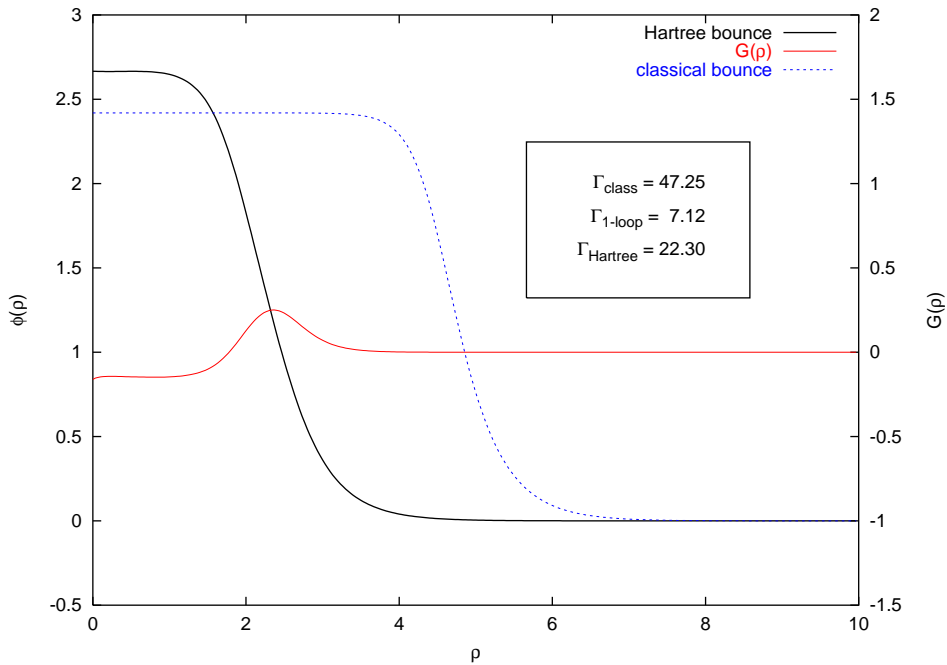} 
\caption{\label{scbfig2}  The self-consistent bounce and coincident 
two-point function for the $\phi^6$ model with parameters 
$\lambda/\mu^2 = 0.03, \epsilon/\mu^2=0.15$ are shown in solid 
lines. The dashed line is the (semi)classical bounce for this 
model. (Note that the $y$-axis for $G(\rho)$ is shifted with respect to 
the axis labeled by $\phi(\rho)$; $G$ still goes to zero in the false vacuum.) 
Inset: Euclidean action for various approximations.}  
\end{figure} 
 
\begin{figure} 
\vspace{0.5in} 
\includegraphics[width=5in]{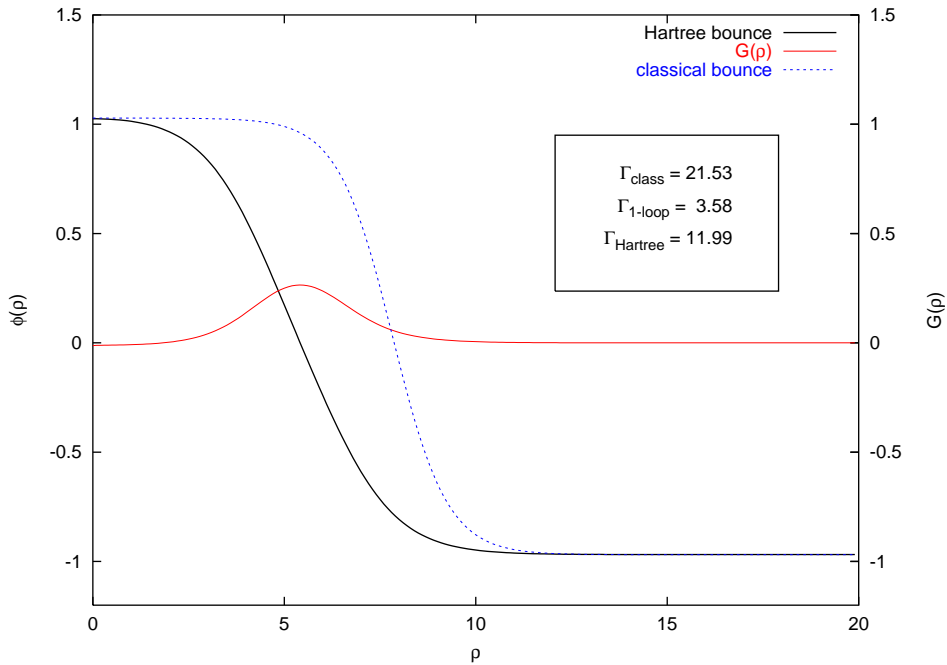} 
\caption{\label{scbfig3} The self-consistent bounce and coincident 
two-point function for the $\phi^4$ model with parameters 
$\lambda/\mu^2 = 1, \epsilon/\mu^2=0.12$ are shown in solid lines. 
The dashed line is the (semi)classical bounce for this model.  Inset: 
Euclidean action for various approximations.}  
\end{figure} 
 
In Figs.~\ref{scbfig2}-\ref{scbfig3} we show the classical bounce along with the 
self-consistent bounce and the coincident two-point function for our 
two models.  In the inset of each we give the value of the Euclidean action at the 
different orders of approximation.  As the figures indicate, it is 
possible for the self-consistent bounce profile to differ 
significantly, both in radius and thickness, from the semi-classical 
prediction.  Moreover the action, which will enter into our rate calculation, 
is dramatically different. 
 
\begin{figure} 
\vspace{0.5in} 
\includegraphics[width=5in]{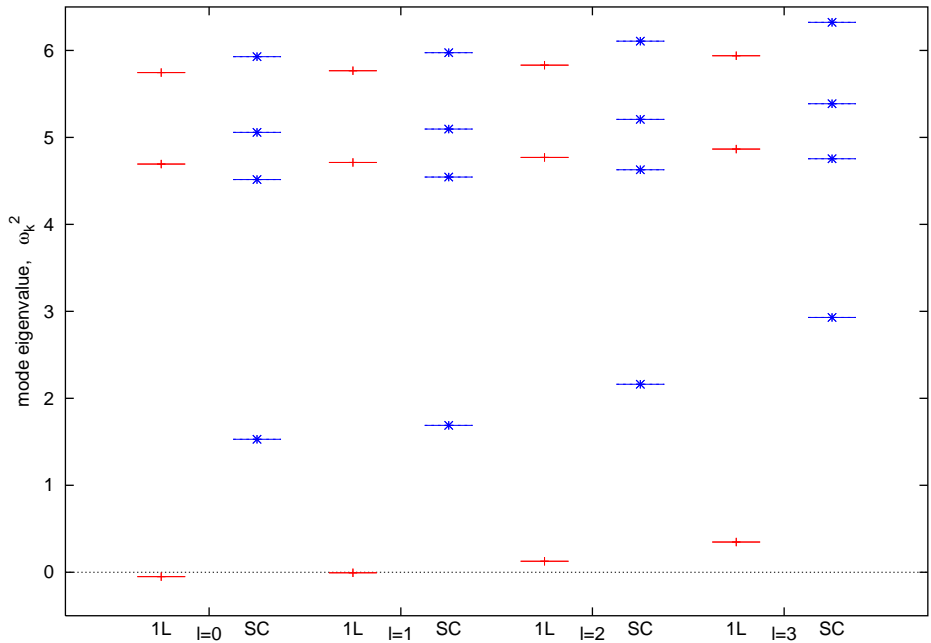} 
\caption{\label{scbspectra} Spectra of the low lying fluctuation 
modes for the $\phi^6$ model.  Plotted are the first few modes in the 
discrete spectrum for the first few partial waves, labeled by the 
quantum number $l$.  The 1-loop spectra are shown side-by-side with 
the self-consistent spectra, with labels on the 
$x$-axis indicating 1L for 1-loop and SC for self-consistent. } 
\end{figure} 
 
It is important to mention also the difference between the 
self-consistent fluctuation spectrum and the 1-loop fluctuation 
spectrum.  We display a plot of the low-lying spectra for the 
$\phi^6$ model in 
Fig.~\ref{scbspectra}.  Modes are labeled by radial and angular quantum 
numbers $n$ and $l$ in Eq.(\ref{eigenvalue}), and in the figure the $x$-axis 
proceeds along the increments in the angular number $l$. The 1-loop 
and self-consistent spectra for each $l$ are shown side-by-side for comparison. 
The numerical spectrum computed in a finite volume is of course 
discrete, but the lowest four states shown are actually bound states. 
 
What is noteworthy is that the negative mode and the two zero modes of 
the 1-loop spectrum (each mode for $l \ne 0$ is doubly degenerate) are 
no longer negative or zero in the self-consistent spectrum.  In fact, 
this is hardly surprising: once the fluctuation modes are allowed to 
interact in the presence of the bounce background, the spectrum will of 
course be shifted to positive values as the modes stabilize each 
other \cite{Boyanovsky, kinkpaper}.  Needless to say this does  
not mean that translational invariance of the center of the bounce has 
been lost.  Translational invariance is however no longer manifest at the 
level of linearized fluctuations since the mean field now cannot be 
translated without also translating the coincident two-point function 
$G$.  We will return to the issues raised by this fact in the next 
section.  For now it is perhaps most clearly illustrated at the level of the equations of 
motion.   
 
Consider for example,  the $\phi^4$ model.  At the 1-loop level, we 
start with the usual equation of motion for the field and for 
linearized fluctuations around the bounce.  
\beqar 
&& \Box \phi + V'(\phi) = 0 \\ 
&& \left[\Box + V''(\phi)\right] \psi_k(x) = \omega^2_k \psi_k(x) 
\eeqar 
It is evident by taking the derivative of the field equation that each 
of the modes corresponding to translation of the field $\partial_{\vec x} 
\phi$ satisfy the fluctuation equation with zero eigenvalue.  A 
similar trick will not work on the set of self-consistent equations in 
(\ref{phieq4}) and (\ref{Geq4}), which we recast below as follows: 
\beqar 
&&\Box \phi + V'(\phi) + 3 \lambda \phi G(x,x) = 0 \\ 
&& \left[\Box +V''(\phi) +3\lambda G(x,x)\right] \psi_k(x) = \omega^2_k\psi_k(x)  
\eeqar  
The coincident two-point function $G(x,x)$ now acts as an effective 
source in the fluctuation equation. 
\section{Collective coordinates and the nucleation rate} 
\label{newrate}
  
As we have just seen the low lying fluctuation spectrum in the 
self-consistent case is qualitatively different from the spectrum at 
1-loop. The absence of the negative and zero modes is a direct 
consequence of the mode self-consistency and now implies that 
translational invariance is no longer a symmetry of the { \it mean-field }
bounce profile.  Nevertheless it clearly remains a symmetry of the 
transition amplitude. This situation is reminiscent of 1-loop  
calculations where the path integral possesses more (e.g. internal) 
symmetry than exhibited by the bounce solution 
\cite{Buchmuller,Kusenko,KusenkoWeinberg}. In these cases the 
contributions from such degrees of freedom must be accounted for in 
the transition amplitude (and consequently the rate of decay) via 
collective coordinates. That is the goal of the present section.  
 
The existence of one negative eigenmode and $D$ zero eigenmodes in the 1-loop case carries   
important physical information about the nature of the bounce solution and its symmetries.   
The former indicates that there is an instability related with the 
bubble's expansion or contraction, whereas the latter are a 
consequence of the translational invariance of the bubble as a 
whole. In 1+1 dimensions these modes are related as they refer to displacements of the 
bubble  walls in opposite directions or in the same direction, 
respectively. The translational and dilatational excitations  
can in fact be built up from the zero modes of the bubble walls as 
their symmetric and their anti-symmetric combinations in the thin wall
approximation \cite{CarvalhoBoyanovsky}. 
  
\begin{figure}  
\vspace{0.5in}  
\includegraphics[width=5in]{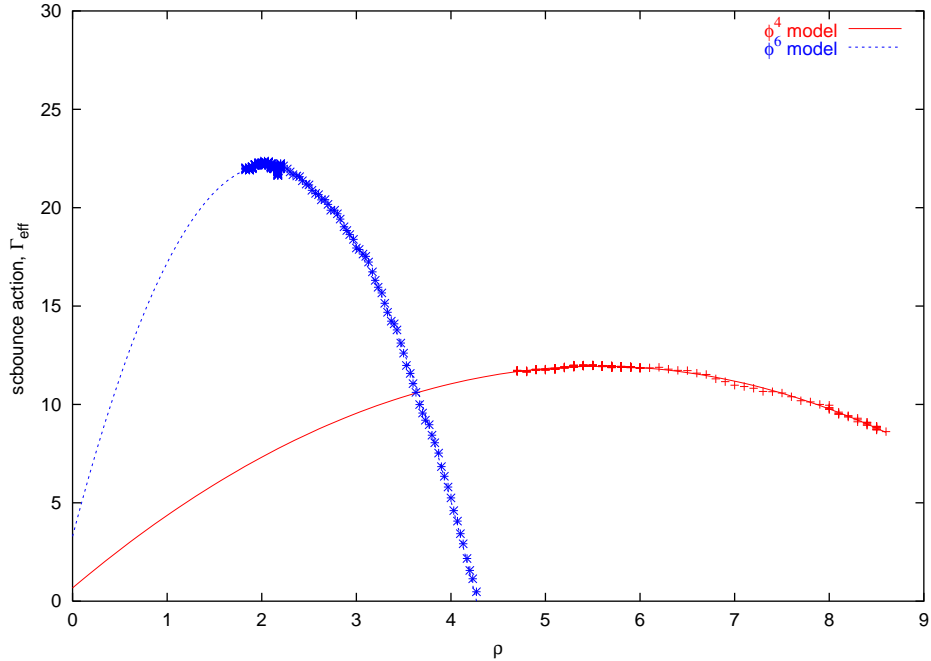}  
\caption{\label{actionvsradius} The self-consistent bounce action for  
both models is plotted as a function of the bounce radius as the  
bounce is dilated and contracted.  The curve fits are quadratic functions and require  
only the location of the peak and the second derivative (curvature) as parameters.}   
\end{figure}  
  
In the self-consistent case the bubble growth instability and its 
translational invariance persist globally in the effective action, but 
are no longer manifest at the level of the fluctuation spectrum, which 
now consists  solely of positive, ``vibrational'' modes. As we have 
alluded, translational invariance now requires simultaneous 
displacements of both $\phi(x)\ra \phi(x+a)$ and $G(x,x) \ra 
G(x+a,x+a)$. Similarly the bubble instability--the fact that it is a
saddle point of the effective action--can be probed 
via a simultaneous radial dilation or contraction of both $\phi$ and 
$G$, see Fig.~\ref{actionvsradius}.   
  
Care must be taken when tracing over the collective coordinates because   
the contributions from these degrees of freedom carry dimensions and potentially lead 
to new (infrared) divergences.  The regularization of such behavior should not introduce 
new {\em ad hoc} scales in the calculation but rather should be consistent with the regularization and renormalization implemented in Sec.~\ref{screnorm}. 
In practice, the dimensionality introduced by the collective coordinates is compensated for 
in the transition amplitude by an infrared regulator $\mu_0$, which is
tied to the choice of renormalization scale made above in    
dealing with the logarithmic divergences in the effective action \cite{tHooft}.  The
physical (measured) mass of quasiparticles is a function of this scale
$M=M(\mu_0)$.  Thus by fixing the mass in the 
false vacuum, we have implicitly adopted the scale choice
$\mu_0=M_+$ \footnote{Another possible scale is given by the inverse object radius,
i.e. $\mu_0=R_0^{-1}$ \cite{tHooft}. As can be shown explicitly in the
thin wall approximation, this latter choice ties the regulator scale
to the symmetry breaking parameter $\epsilon$.}.  


We now evaluate the (consistently regularized) contribution from  the collective 
coordinates, beginning with the translational zero modes.  The
effective action is invariant under a transformation which shifts the
center of the   bounce. As usual, the path integral obtains a factor
of spacetime volume up to a  
Jacobian factor $J_0$, which in $D$ dimensions is given by 
\begin{eqnarray}  
D[\phi(x,z)] = J_0 d^D z, \qquad {\rm with} \quad J_{0} = \left[{1 
\over 2 \pi}  \int d^D x \left( {\partial \phi_b \over \partial z} \right)^{2} \right]^{D/2}. 
\end{eqnarray}  
where the factor of $2 \pi$ is conventional to make the integral commensurate with  
a the Gaussian case.   
  
The integral can be expressed in terms of the effective action. To be
explicit, we wish to consider the effective action now as a functional
of $\phi$ alone.  This amounts to the familiar transformation of the
2PI  action $\Gamma_{\rm eff} [\phi,G]$  into the 1PI
effective action $\Gamma^{\rm 1PI}_{\rm eff} [\phi]$, wherein $G$ is thought of
as a functional of $\phi$ determined by Eq.~(\ref{propeq}) \cite{2PI}.
Dividing this action into kinetic and effective potential terms, the
$\phi$ equation is simply
\begin{eqnarray}  
-\nabla^2 \phi  + {\delta V_{\rm eff} \over \delta \phi} = 0 . 
\end{eqnarray}  
and we have essentially the same situation as in the 1-loop analysis
but for the substitution of an effective potential.  The argument
proceeds that the effective action is conserved along a trajectory in $x$:    
$\partial_x \Gamma_{\rm eff}=0$ and its value is fixed by our choice  
of renormalization in the false vacuum $\Gamma_{\rm eff} [\phi_+] = 0.$  
As a consequence  we can write   
\begin{eqnarray} 
{1 \over 2} \left ( {d\phi \over dx} \right)^2  =   V_{\rm eff}, \quad {\rm and} \quad  
J_0 = \left[ {\Gamma_{\rm eff} \over 2 \pi} \right]^{D/2}.  
\end{eqnarray}  
The total contribution from the zero modes, together with factors of
the regulator $\mu_0$, is   
\begin{eqnarray} 
J_{0}  \mu_0^D \int d^{D} z.   
\label{zeromodecontribution}  
\end{eqnarray}  
  
Next we consider what happens upon variation of the radius 
corresponding to the bounce solution in $\phi$ and $G$. It is 
convenient to shift the radial coordinate by the bounce radius,
i.e. $r= R_{0} + \xi$, where $r=R_{0}$  is the radius of the  
bounce.  Expanding the effective action around $\xi=0$ then gives 
\begin{eqnarray}  
\Gamma_{\rm eff} \simeq \Gamma_{\rm eff} (R_0) + \xi \Gamma'_{\rm eff}  (R_0)   
+ {1\over 2}  \xi^2 \Gamma''_{\rm eff}  (R_0)  + \ldots  
\end{eqnarray}  
where primes denote radial derivatives. 
The first derivative of the effective action at the bounce is zero as 
the bounce is a solution of the equations of motion. The second derivative  
however is negative, as shown in Fig.~\ref{actionvsradius}.  Just as 
is the case for the semiclassical solution, the self-consistent
bounce is a saddle point of the effective action and a maximum in the
radial ``direction'' (by which we mean direction in function
space). The contribution of this dilatational degree of 
freedom to $\cal Z$ is computed  
in direct analogy to that of the negative mode in the
1-loop analysis, i.e. in steepest descent by continuation to imaginary $\xi$ 
\cite{CallanColeman}. The difference is that we must now extract $\Gamma''_{\rm eff}$ 
by hand, so to speak, rather than from the fluctuation spectrum.  Performing the integration 
in the radial coordinate requires again the introduction of a Jacobian  
\begin{eqnarray}  
J_{-} =  \left[{1 \over 2 \pi}  \int d^Dx \left( {\partial \phi_b \over \partial R} \right)^{2} \right]^{1/2}.  
\end{eqnarray}  
The change to radial coordinates has no effect, and $J_-$ is identical 
to $J_0$ for one dimension. 
 
The $D+1$ collective coordinates give a total multiplicative contribution to $\cal Z$ of 
\begin{eqnarray}  
 {\mu_0}^{D+1}\left( {\Gamma_{\rm eff} \over 2 \pi} \right)^{D+1 \over 2}   \int d^D z \int d \xi ~   
 e^{- \Gamma''_{\rm eff} \xi^2/2} ,
\end{eqnarray}  
or, performing the Gaussian integration,  
\begin{eqnarray}  
\left( {\Gamma_{\rm eff} \over 2 \pi} \right)^{D+1 \over 2} {i
\mu_0^{D+1} \over 2} \sqrt{ {2 \pi \over  \Gamma''_{\rm eff}}}   \int
d^D z ,
 \end{eqnarray}  
where the factor of $1/2$ results from the analytic continuation to one 
 branch of imaginary $\xi$ since $\Gamma''_{\rm eff}$ is negative.   
  
 
Finally the rate of spontaneous decay of the false vacuum is obtained 
by Eq.~(\ref{imagpart}),  
\begin{eqnarray}  
\Gamma/V =  {\mu_0}^{D+1} \left[ {\Gamma_{\rm eff} \over 2 \pi}\right]^{D+1 \over 2} \left( 2 \pi \over \Gamma''_{\rm eff} \right)^{1/2}  ~e^{- \Gamma{\rm eff}},  
\label{rate}
\end{eqnarray}  
where all factors of  $\Gamma_{\rm eff}$ and its derivatives are 
evaluated at the self-consistent bounce.  The dimensions are correct 
as the factor containing derivatives of the effective action cancels
one mass dimension.  We present numerical results for the 
nucleation rate computed according to our self-consistent method 
described above and computed at 1-loop order in
Table~\ref{nucrates}. The models are as in Eqs.(\ref{p4model})-(\ref{p6model}) with parameters
given by the numerical values $\mu^2=4.34, \lambda = 0.14, \epsilon=0.67$  for the $\phi^6$
model and $\lambda = mu^2 = 1, \epsilon=0.12$ for the $\phi^4$
model.
 
\begin{center} 
\begin{table} 
\begin{tabular}{lcccc}  
\hline \hline 
 && $\Gamma_{\rm sc}/V$ & & $\Gamma_{\rm 1-loop}/V$  \\ 
\hline 
$\phi^4$ model & & $1.19 \times 10^{-4}$ & & $8.59 \times 10^{-3}$ \\ 
$\phi^6$ model & & $1.23 \times 10^{-8}$ & & $1.60 \times 10^{-2}$  \\ 
\hline \hline 
\end{tabular} 
\caption{Numerical results for the self-consistent nucleation rate 
and the 1-loop rate in the two scalar models.} 
\label{nucrates} 
\end{table} 
\end{center} 

\section{Discussion}
 \label{discussion}

We see that the self-consistent calculation  results in a much smaller nucleation rate than 
the usual 1-loop estimate, from two orders of magnitude in the
$\phi^4$ model up to six orders of magnitude for the $\phi^6$
model. This is a robust prediction even if the nonexponential
prefactor in Eq.~(\ref{rate}) is taken with a grain of salt.  The
prefactor is of the same order of magnitude in both the
self-consistent and 1-loop approximations, hence it is the
exponentiation of the effective action which mostly determines the magnitude
of the rate.  We should clarify that the ``dynamical'' prefactor,
i.e. the ratio of fluctuation determinants in Eq.~(\ref{1loopRate}), is not generically of
order unity.  It is only the case that when the (appropriately dimensionless)
determinant contribution is absorbed into the effective action in the exponent that
it may be argued that the remaining prefactor is roughly determined by some mass
scale typical of the problem as $M^D$.

The larger effective action at the self-consistent level results
principally from the the shift in the spectrum (relative to
1-loop) by the fluctuation self-repulsion. The same qualitative change
is observed for self-consistent topological defects, which are heavier
than at 1-loop \cite{kinkpaper}, and we expect it to be generically
true for any self-consistently dressed quantum field configuration.
 
In summary, we have proposed an alternative nucleation rate computation based on a
self-consistent determination of the bubble shape and fluctuation
 spectrum.  We have traced a path analogous to the standard work of
 Coleman \cite{Coleman} by going to imaginary time and solving instead
 a set of equations for  the mean-field bounce and coincident
 two-point function obtained from a 2PI effective action.  As
 translational invariance and bubble instability are not manifest at
 the level of the linearized fluctuation spectrum, we have
 used collective coordinates to account for the contribution of these
 degrees of freedom to the transition amplitude.

Finally  we note that as our Euclidean spacetime solution, upon analytically continuation to
 Minkowski spacetime should provide a solution to the real time bubble
 dynamics {\em for all time} in accordance with Lorentz invariance at zero temperature
 \cite{stepbeyond}.  This is the ultimate test of whether or not what
 we claim above to be the self-consistent bounce (at this level of
 approximation) {\em is} indeed the self-consistent bounce. 

Analytic continuation of the one- and two-point functions $\phi(\rho)$ and 
$G(\rho)$ from the last section is a straightforward matter of taking $\rho \ra \sqrt{x^2-t^2}$, or at
initial time $t=0$, $\rho \ra |x|$.  This is not however enough information to specify the real time problem.  Time evolution of the coupled system of equations requires full knowledge of the spectral decomposition of $G(x,x')$.   In order to initialize these, we need to rebuild in {\em
real} time the fluctuation modes that correspond to the (initially
static) $\phi(x)$ and $G(x,x)$. In practice we use the methods developed in Ref.~\cite{kinkpaper}
to easily determine the mode functions that correspond to $\phi(x)$ and $G(x,x)$ found 
in the previous sections. We then use them as initial conditions for the real time field evolution.

\begin{figure}
\begin{center}
\includegraphics[width=5in]{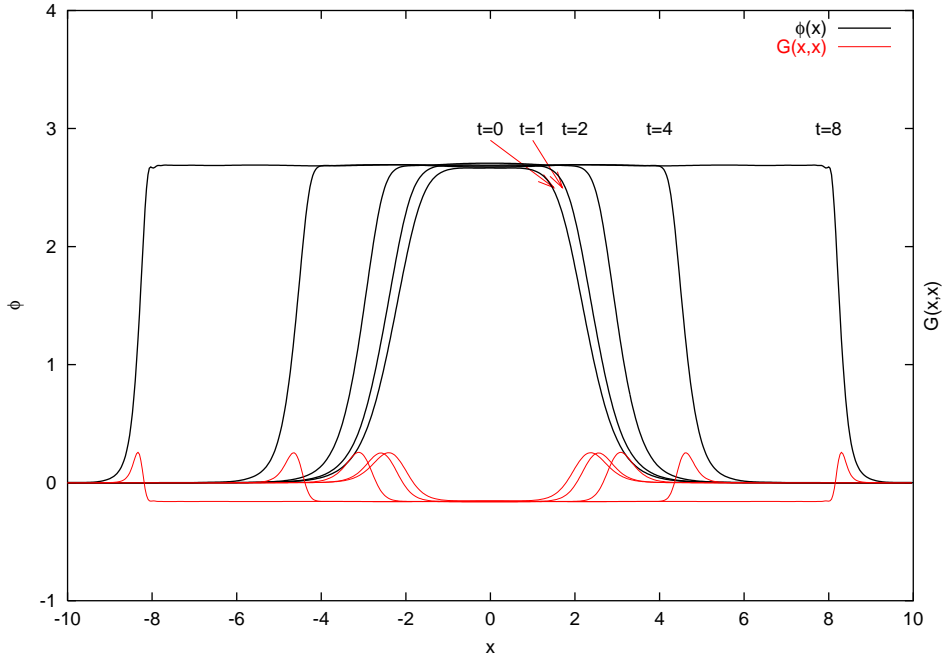}
\caption{A stroboscopic view of the bubble growth for the
  self-consistent bubble of the $\phi^6$ model. The bubbles evolve by 
expansion of the true vacuum and Lorentz contraction of the walls, without giving rise to radiation, as required of the self-consistent bounce.}
\label{p6rt}
\end{center}
\end{figure}

\begin{figure}
\begin{center}
\includegraphics[width=5in]{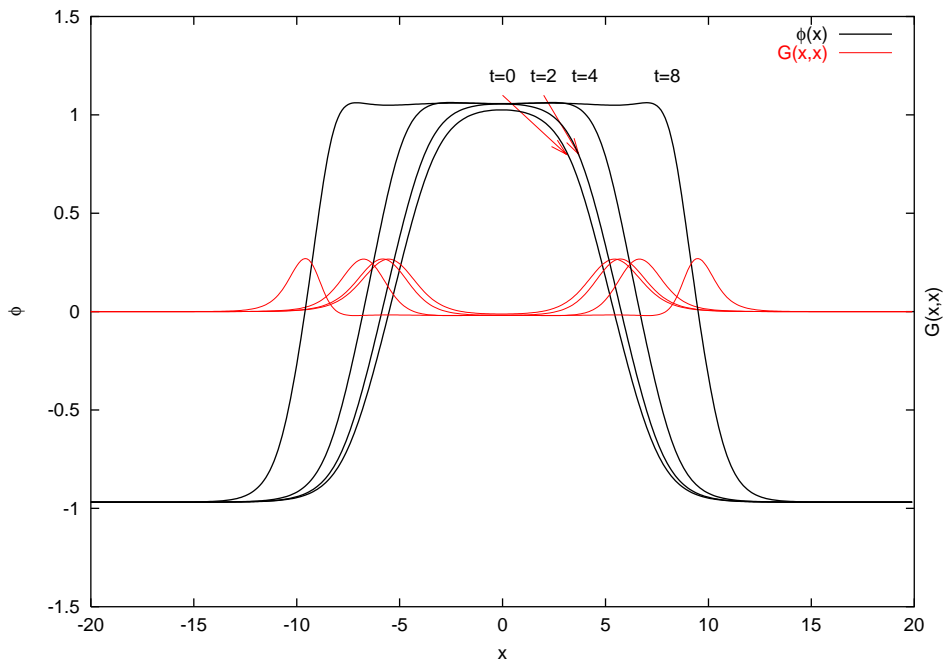}
\caption{The same as Fig.~\ref{p6rt} for the $\phi^4$ model.}
\label{p4rt}
\end{center}
\end{figure}

At zero temperature, we should therefore observe a Lorentz invariant bubble
growth without the initial transients present in the naive initializations  of 
Ref.~\cite{stepbeyond}.  
As figures \ref{p6rt} and \ref{p4rt}, corresponding to the $\phi^6$ and $\phi^4$ models, indicate 
the shape of the bubble is essentially unaltered during the evolution, except for Lorentz contraction.

\end{document}